\documentclass[a4paper,fleqn,usenatbib]{mnras}

\usepackage{newtxtext,newtxmath}

\usepackage[T1]{fontenc}
\usepackage{ae,aecompl}

\usepackage{graphicx}	
\usepackage{amsmath}	
\usepackage{pdflscape}

\title[NGTS Observations of DW Cnc]{The return of the spin period in DW Cnc and evidence of new high state outbursts}

\author[C. Duffy et al.]{C. Duffy,$^{1,2}$\thanks{Contact e-mail: \href{mailto:christopher.duffy@armagh.ac.uk}{christopher.duffy@armagh.ac.uk}}
G. Ramsay,$^{1}$
D. Steeghs,$^{2,8}$
M. R. Kennedy,$^{3,4}$
R. G. West,$^{2}$
P. J. Wheatley,$^{2}$
V. S. Dhillon,$^{5, 6}$
\newauthor
K. Ackley,$^{2,7,8}$
M. J. Dyer,$^{5}$
D. K. Galloway,$^{7,8,9}$
S. Gill,$^{2}$
J.~S.~Acton,$^{10}$
M.~R.~Burleigh,$^{10}$
S.~L.~Casewell,$^{10}$
\newauthor
M.~R.~Goad,$^{10}$
B.~A.~Henderson,$^{10}$
R.~H.~Tilbrook,$^{10}$
P.~A.~Str\o{}m,$^{2}$
D.~R.~Anderson $^{2}$
\\ \\
$^{1}$Armagh Observatory and Planetarium, College Hill, Armagh, BT61 9DB, UK\\
$^{2}$Department of Physics, University of Warwick, Gibbet Hill Road, Coventry, CV4 7AL, UK\\
$^{3}$Department of Physics, University College Cork, College Road, Cork, Ireland\\
$^{4}$Jodrell Bank Centre for Astrophysics, Department of Physics and Astronomy, The University of Manchester, Manchester, M13 9PL, UK\\
$^{5}$Department of Physics and Astronomy, University of Sheffield, Sheffield, S3 7RH, UK\\
$^{6}$Instituto de Astrof\'{i}sica de Canarias, E-38205 La Laguna, Tenerife, Spain\\
$^{7}$School of Physics \& Astronomy, Monash University, Clayton VIC 3800, Australia\\
$^{8}$OzGrav: The ARC Centre of Excellence for Gravitational Wave Discovery, Clayton VIC 3800, Australia\\
$^{9}$Institute for Globally Distributed Open Research and Education (IGDORE) \\
$^{10}$School of Physics and Astronomy, University of Leicester, University Road, Leicester, LE1 7RH, UK\\
}

\date{Accepted 2021 November 19. Received 2021 November 19; in original form 2021 October 19}

\pubyear{2021}

\begin{document}
\outer\def\gtae {$\buildrel {\lower3pt\hbox{$>$}} \over 
{\lower2pt\hbox{$\sim$}} $}
\outer\def\ltae {$\buildrel {\lower3pt\hbox{$<$}} \over 
{\lower2pt\hbox{$\sim$}} $}

\label{firstpage}
\pagerange{\pageref{firstpage}--\pageref{lastpage}}
\maketitle

\begin{abstract}
DW Cnc is an intermediate polar which has previously been observed in both high and low states. Observations of the high state of DW Cnc have previously revealed a spin period at $\sim38.6$ min, however observations from the 2018/19 low state showed no evidence of the spin period. We present results from our analysis of 12 s cadence photometric data collected by NGTS of DW Cnc during the high state which began in 2019. Following the previously reported suppression of the spin period signal we identify the return of this signal during the high state, consistent with previous observations of it. We identify this as the restarting of accretion during the high state. We further identified three short outbursts lasting $\sim1$ d in DW Cnc with a mean recurrence time of $\sim 60$ d and an amplitude of $\sim1$ mag. These are the first outbursts identified in DW Cnc since 2008. Due to the short nature of these events we identify them not as a result of accretion instabilities but instead either from instabilities originating from the interaction of the magnetorotational instability in the accretion disc and the magnetic field generated by the white dwarf or the result of magnetic gating.
\end{abstract}

\begin{keywords}
accretion, accretion discs -- binaries: close -- novae, cataclysmic variables -- instabilities -- stars: individual: DW Cnc
\end{keywords}



\section{Introduction}

\begin{figure*}
\includegraphics[width=\textwidth]{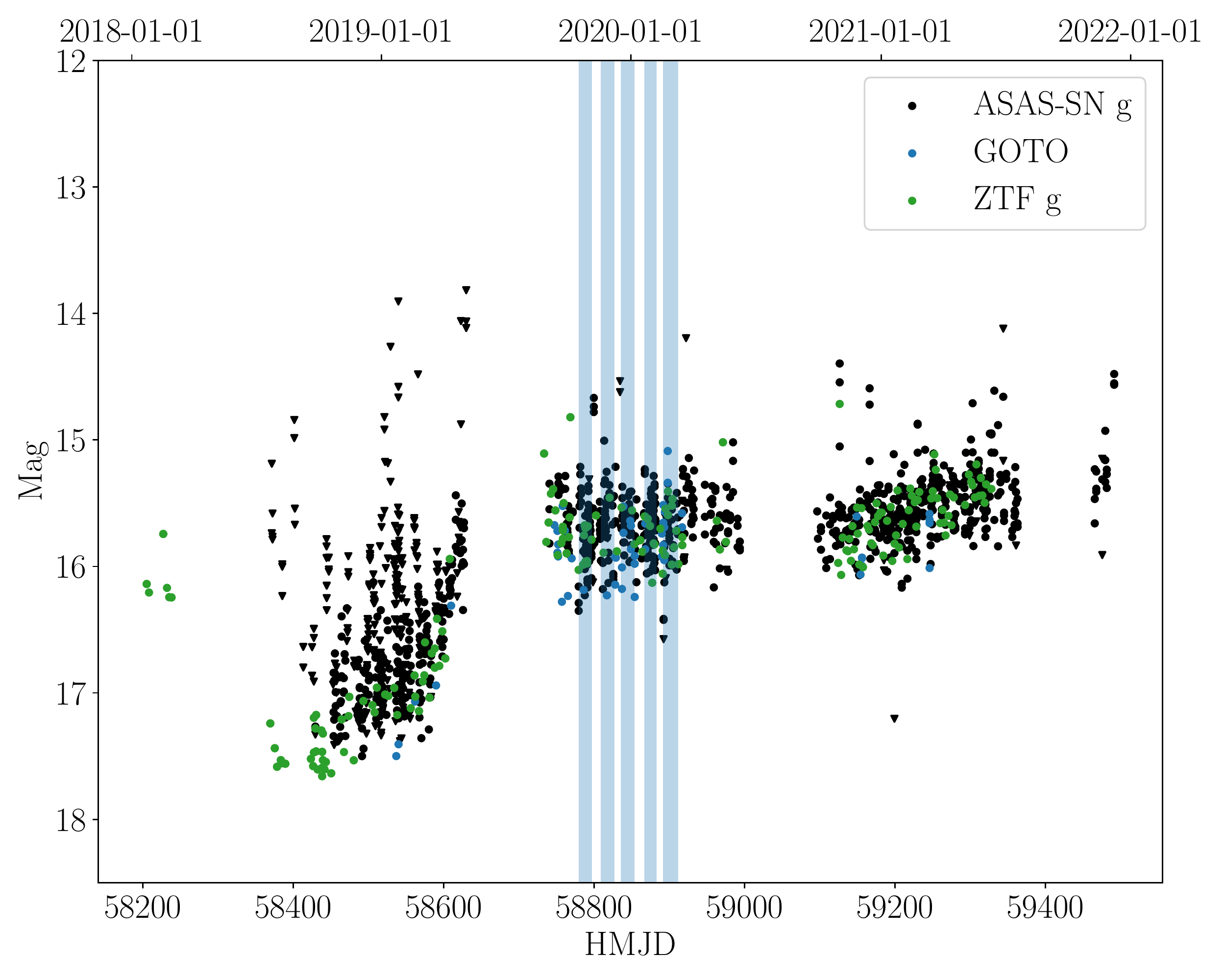}
\caption{A lightcurve of DW Cnc spanning approximately 4 years using data obtained from ZTF and ASAS-SN in the \textit{g} band and GOTO in the \textit{L} band. The triangular markers denote measurements which are upper limits. The blue vertical lines indicate the times of the observations made using NGTS which are shown in \autoref{NGTS_curve}.}
\label{All_Sky}
\end{figure*}

Cataclysmic variables (CV) are systems composed of a white dwarf (WD) primary and a late type secondary, making up an interacting binary system, where the secondary star overflows its Roche Lobe resulting in accretion from the secondary on to the WD. In most systems this accretion results in the formation of a disc around the WD before the material accretes onto the photosphere of the WD. However, if the WD has a magnetic strength field \gtae1MG, an accretion disc is truncated or does not form and the accretion flow follows the magnetic field lines onto the magnetic poles of the WD, such systems are known as ``Polars''. In Polars the rotational period of the WD is equal to the orbital period of the system, the accretion stream is therefore ``locked'' into following the magnetic field lines.

However in WDs which have a magnetic field between $\sim$0.1-10 MG, known as ``Intermedidate Polars'' (IPs) \citep[see][for an early review]{patterson1994}, an accretion disc can be formed. The accretion disc can however become truncated by the magnetic field which results in some of the accreting material overflowing the disc and accreting directly along the field lines as a stream \citep[see][]{1996ApJ...470.1024A}. In these IPs the rotation period of the WD is not equal to the orbital period and the ratio of the rotation period of the WD to the binary orbital period is typically $\sim$0.01--0.6 \citep[e.g.][]{norton2008}. The accretion flow forms a bright spot at the magnetic pole; if the magnetic axis and the spin axis are not aligned then this bright spot appears to sweep around in what is known as the lighthouse effect. This effect is synchronous with the spin period and is useful in revealing it to us \citep{warner2003cataclysmic}.

A number of IPs have been observed to outburst previously, often with long recurrence times between outbursts which are themselves short lived, sometimes only a matter of hours, e.g. $\sim6$ hours in TV Col \citep{1984ApJ...280..729S} and $\sim6-24$ hours in V1223 Sgr \citep{1989A&A...219..195V}. These outbursts have been characterised similarly to those from Dwarf Novae, and described with success using the Disk Instability Model (DIM) \citep{1997MNRAS.292..397H}. More recently \citet{2017A&A...602A.102H} modelled the outbursts from IPs concluding that outbursts from those systems that last a few or more days could indeed by described successfully using the DIM but that the shortest events on the order of hours could not. They suggested that these events were more likely the result of the interaction between the magnetic field generated by the magnetorotational instability of the accretion disc and the WD magnetic field resulting in a short lived increase in the mass transfer rate.

Short outbursts of duration $\sim1-2$ h have been identified in TW Pic by \citet{2021NatAs.tmp..201S} and it is thought that these events are the result of magnetic gating. Magnetically gated outbursts can occur where the co-rotation radius (where the Keplerian frequency of the accretion disc and the WD spin frequency are equal) and the magnetospheric radius are similar, creating a barrier to accretion. As a result of this barrier material builds up in the disc until some critical value at which point the barrier is overcome in a short lived accretion event \citep{2010MNRAS.406.1208D,2012MNRAS.420..416D}.

DW Cnc was first identified as a variable star by \citet{1982PZ.....21..691S} and as a  magnetic CV by \citet{2004MNRAS.349..367R}. DW Cnc has two distinct states; a high and a low state -- with V band magnitudes of $\sim15$ and $\sim17.5$ respectively \citep{1982PZ.....21..691S}. In addition to high states DW Cnc occasionally shows outbursts, the first seen by \citet{2008JAVSO..36...60C} who observed a brightening to a peak of V$\sim11.4$ mag in an outburst that lasted $<$7 days. 

Observations of DW Cnc have revealed a number of periodic signals of various origins. A period of 86.1 min, attributed to the orbital period ($P_{orb}$), and a period of 38.6 min, attributed to the spin period of the WD ($P_{spin}$), was identified by \citet{2004MNRAS.349..367R} and \citet{2004PASP..116..516P} respectively. Beat periods, which are interactions of the signal of $P_{orb}$ and $P_{spin}$,  have also been seen in DW Cnc  with \citet{2004PASP..116..516P} identifying one at 69.9 min and another by \citet{2008JAVSO..36...60C} at 73.7 min. In addition, a weak periodic variation at 110.9 min has been detected, although no physical origin has of yet been attributed to this signal \citep{2004PASP..116..516P,2008JAVSO..36...60C}. The spin period has also been identified in X-rays, with an X-ray spectrum being typical of an IP \citep{Nucita2019}. 

\citet{2020MNRAS.494.4110S} using spectroscopic observations found that the $P_{spin}$ signal was much weaker during the low state compared to the high state when $P_{spin}$ is mostly the strongest signal followed by the 69.9 min beat period, followed by $P_{orb}$. \citet{2020MNRAS.494.4110S} argue that the significant weakening of the signal due to $P_{spin}$ is due directly to the system being in a low state and they expected to see the return of the previous paradigm upon the return to a high state. To test this prediction we have used data of DW Cnc taken using the Next Generation Transit Survey \citep[NGTS;][]{2018MNRAS.475.4476W} when the system returned to a high state.

\begin{figure*}
\includegraphics[width=\textwidth]{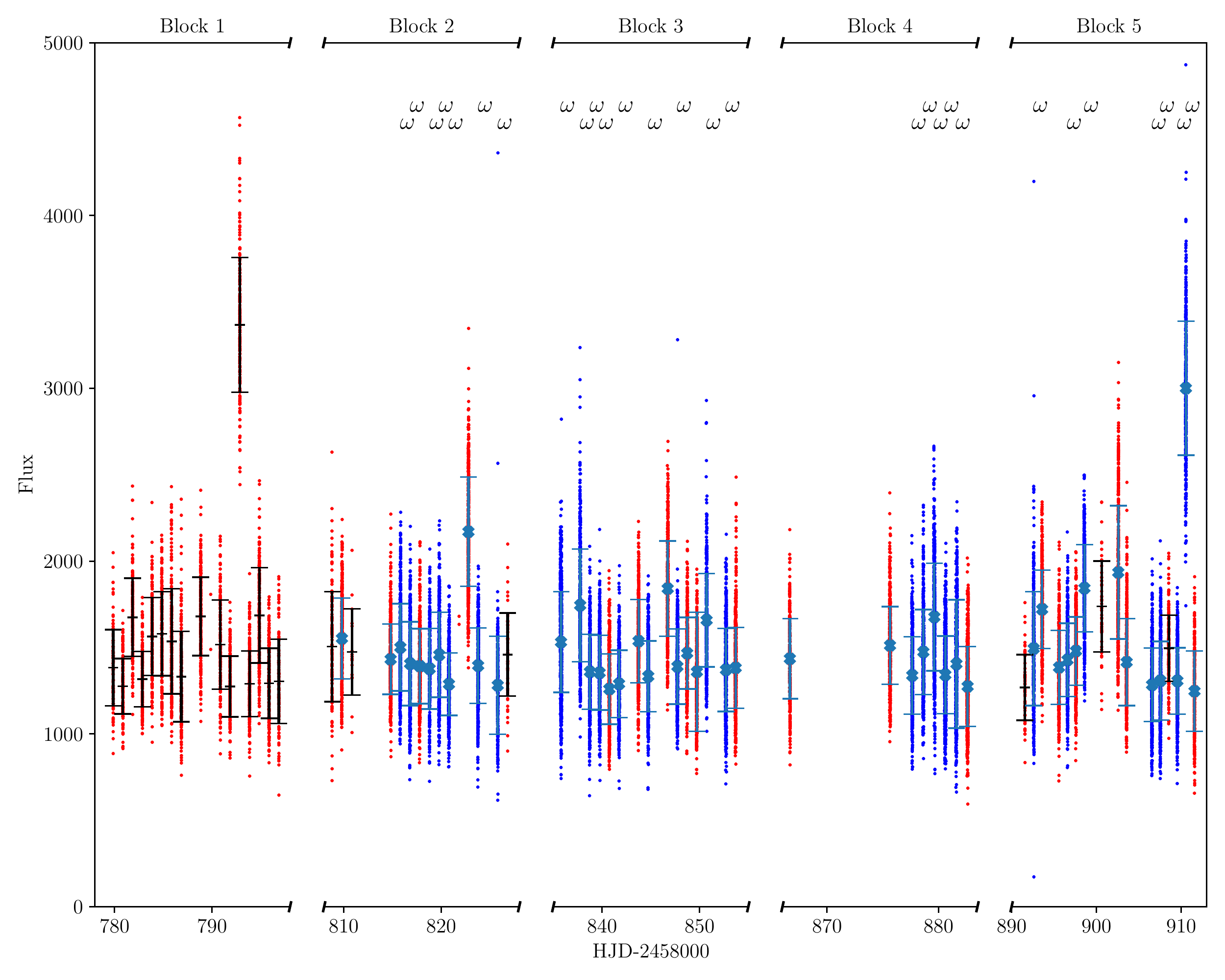}
\caption{Lightcurve of DW Cnc in the five observation blocks covered by NGTS; it has been compressed in time to remove the gaps between observation blocks. A blue $X$ identifies the nightly mean flux, this has been replaced with a black $+$ in nights where the observations are deemed not to have been long enough to search for the spin period; the error bars on each indicate the nightly standard standard deviation. The colour coding denotes which if any periods we were able to identify in a given night. Red denotes no period whilst blue denotes the presence of the spin period, which is also denoted by the presence of $\omega$ above a given night.}
\label{NGTS_curve}
\end{figure*}

\section{Data}

We show the long term light curve of DW Cnc in \autoref{All_Sky} which was made using all sky survey photometry from ZTF \citep{2019PASP..131a8002B}, ASAS-SN \citep{Kochanek_2017} and GOTO \citep{2021arXiv211005539S}. This lightcurve shows that DW Cnc emerged from its low state (which lasted approximately 14 months) back to a high state in May 2019. Additionally this lightcurve shows a number of short brightenings during the high state; most clearly seen from late 2020 onwards. Observations from ZTF indicate that these may be outbursts which peak at a brightness of 13.25 mag in \textit{r} band \citep[private communication]{vanRoestel}.

We have used data taken using NGTS to obtain high cadence photometry over the immediate time interval after DW Cnc had returned to a high state. NGTS is comprised of 12 robotic 20 cm telescopes, with a total instantaneous field of view of 96\degr\textsuperscript{2}, and located at ESO's Paranal Observatory in Chile \citep{2018MNRAS.475.4476W}. Each telescope is fitted with a 2048 × 2048 pixel CCD which has a pixel scale of 5\arcsec/pixel and is sensitive in the range (520-890 nm). The limiting brightness of the standard NGTS pipeline is $\sim$16 mag although it is capable of reliable detections down to 19\textsuperscript{th} mag.

We accessed 12 second cadence photometry of DW Cnc from NGTS as part of its normal observing program. The photometry presented here is reduced via the standard NGTS pipeline which generates photometry using a 3 pixel radius aperture. In addition to this, data associated with high values of excess variance, or airmass were removed in order to exclude photometry which had been adversely effected by these factors.

We accessed five successive observing periods on DW Cnc between 22/10/2019 (HJD=2458779) and 3/3/2020 (HJD=2458912) (noted by blue vertical bands in \autoref{All_Sky}). Each of these periods contained between 9 and 16 nights of observations with each night containing between 40-200 min of continuous photometry. Those nights which do not have data are due in most part to poor weather conditions.  In \autoref{NGTS_curve} we show the lightcurve of each of the five blocks of observations of DW Cnc, compressed to remove the gaps between blocks.

\section{Data Analysis}\label{data_ana}
To search for periodicities in the NGTS data we made use of Lomb-Scargle periodograms as implemented in \texttt{astropy} \citep{2018AJ....156..123T} to identify the periodic behaviour of significance. We also considered an Analysis of Variance \citep{1989MNRAS.241..153S} approach which returned results consistent with the Lomb-Scargle method allowing us to select Lomb-Scargle as our primary method of analysis.

We considered the data on a night by night basis; this was done to avoid complex window functions, arising from regular day or longer gaps in the high cadence observations, which would have had the effect of degrading the period analysis. To determine the minimum duration of data where we would expect to detect $P_{spin}$ and $P_{orb}$, we took lightcurves of various duration from our data and used these to generate synthetic lightcurves with the known spin and orbital periods injected. These injected signals had amplitudes, 0.084 and 0.023 mag respectively, taken from previous findings of \citet{2004PASP..116..516P} with Gaussian noise added to reflect the mean noise in the NGTS data. We found that in order to be able to detect the spin period, observations of at least 60 min were needed. To identify the orbital period 150 min were required although this is heavily dependent on the noise present. As a result of this we only considered lightcurves from nights with $>$60 minutes of observation. In \autoref{NGTS_curve} we note the excluded nights by replacing the cross denoting mean flux with a black $+$. This process excluded 22 nights of observations from our analysis leaving a total of 47 nights with sufficient length to perform a search using the Lomb-Scargle periodogram.

For each of the resulting 47 periodograms we inspected each for the presence of a peak close to the known value of $P_{spin}$ or $2P_{spin}$. A small number of periodograms were rejected at this stage despite having a peak with an origin that was likely to be associated with the spin period; as these peaks were more than 10\% different from the accepted value for the spin period. In total 30 periodograms were identified to contain a spin period peak of interest. We subsequently calculated the significance of those accepted peaks considering those with a significance of at least $3\sigma$ to a contain a reliable signal. The significance of the peak was calculated via a bootstrap method \citep[see][\textsection 7.4.2.3 for details]{2018ApJS..236...16V} for false alarm probability determination at the period of the peak.

\begin{figure}
\includegraphics[width=\columnwidth]{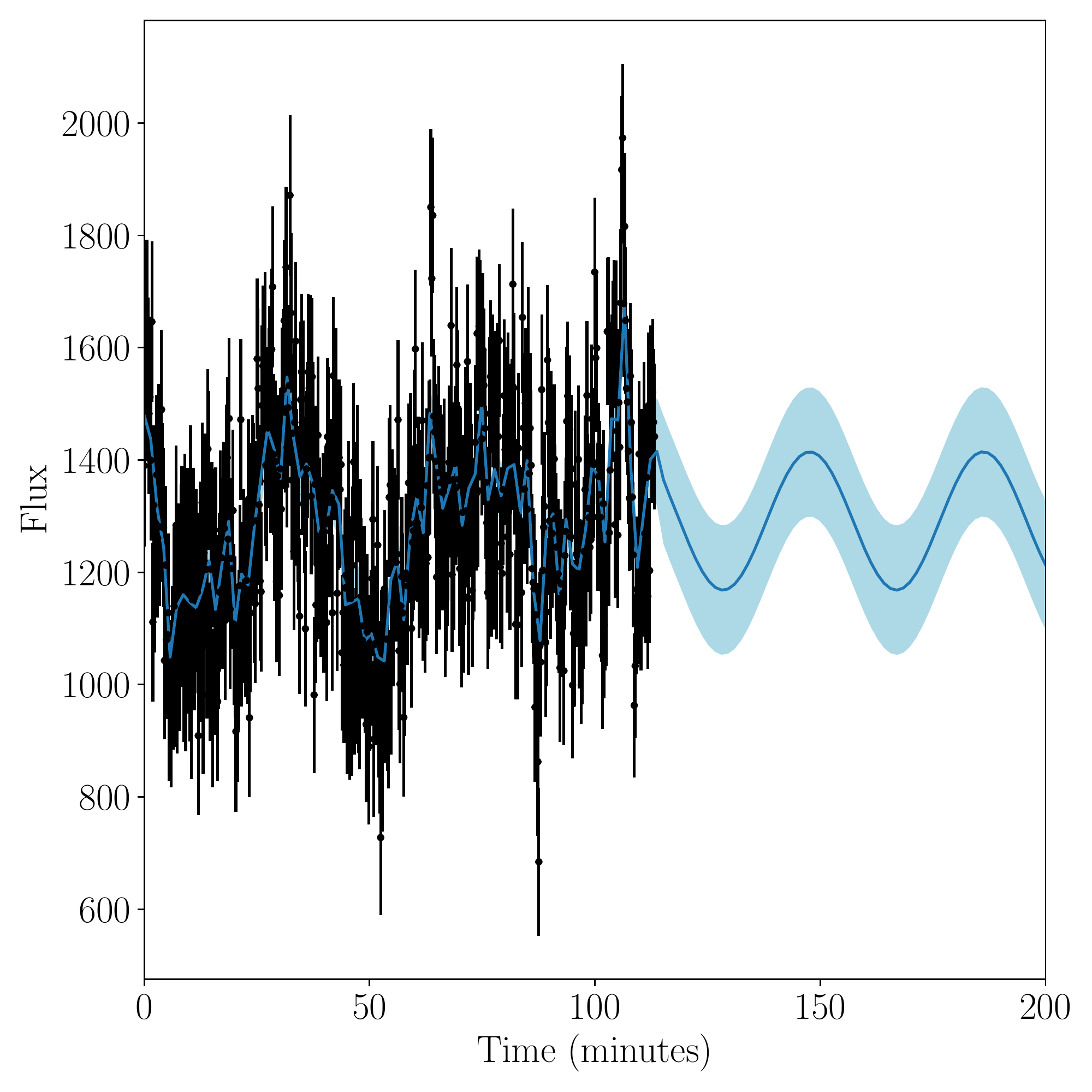}
\caption{Lightcurve, with associated errors, extracted from HJD=2458841 is shown as the black points. The model fited to these data is shown as the blue line with the shaded area denoting the 95\% confidence interval of the fit. The model is extended beyond the data to illustrate the periodic behaviour identified in the model, which shows a clear periodic variation of $\sim40$ minutes.}
\label{model_lc}
\end{figure}

This method is only adequate under the assumption that red noise is not dominant which is not the case in the presence of accretion discs. In order to account for this we confirmed our analysis by using a Gaussian Process regression implemented in \textsc{celerite} \citep{celerite} through a Markov Chain Monte Carlo algorithm \citep{2013PASP..125..306F} to fit a model to the light curve of each night and thus establish the periods present in the data; one such modelled night is shown in \autoref{model_lc}. This analysis relies on modelling the covariance of our data assuming a combination of Gaussian process, represented by different kernels. For these data, we use a combination of a kernel approximating a radial basis function to account for red noise, a periodic kernel to represent the spin period of the white dwarf, and a white noise kernel to represent the measurement uncertainties. The periodic kernel used is that of Equation 56 of \citet{celerite}, and was chosen to allow for efficient sampling of the parameter space when performing the MCMC sampling. For further reading on Gaussian Processes, see \citet{2006gpml.book.....R} and \citet{celerite}.

Through this analysis we were able to confirm that red noise is dominant in these data. Despite this we found that the results of the Lomb-Scargle analysis were consistent with Gaussian Process analysis in terms of the spin period. We therefore have chosen to include the results of Lomb-Scargle analysis it produces periodograms which are easily understood and are entirely consistent with results yielded via other methods.

\section{Results}

\begin{figure*}
\includegraphics[width=\textwidth]{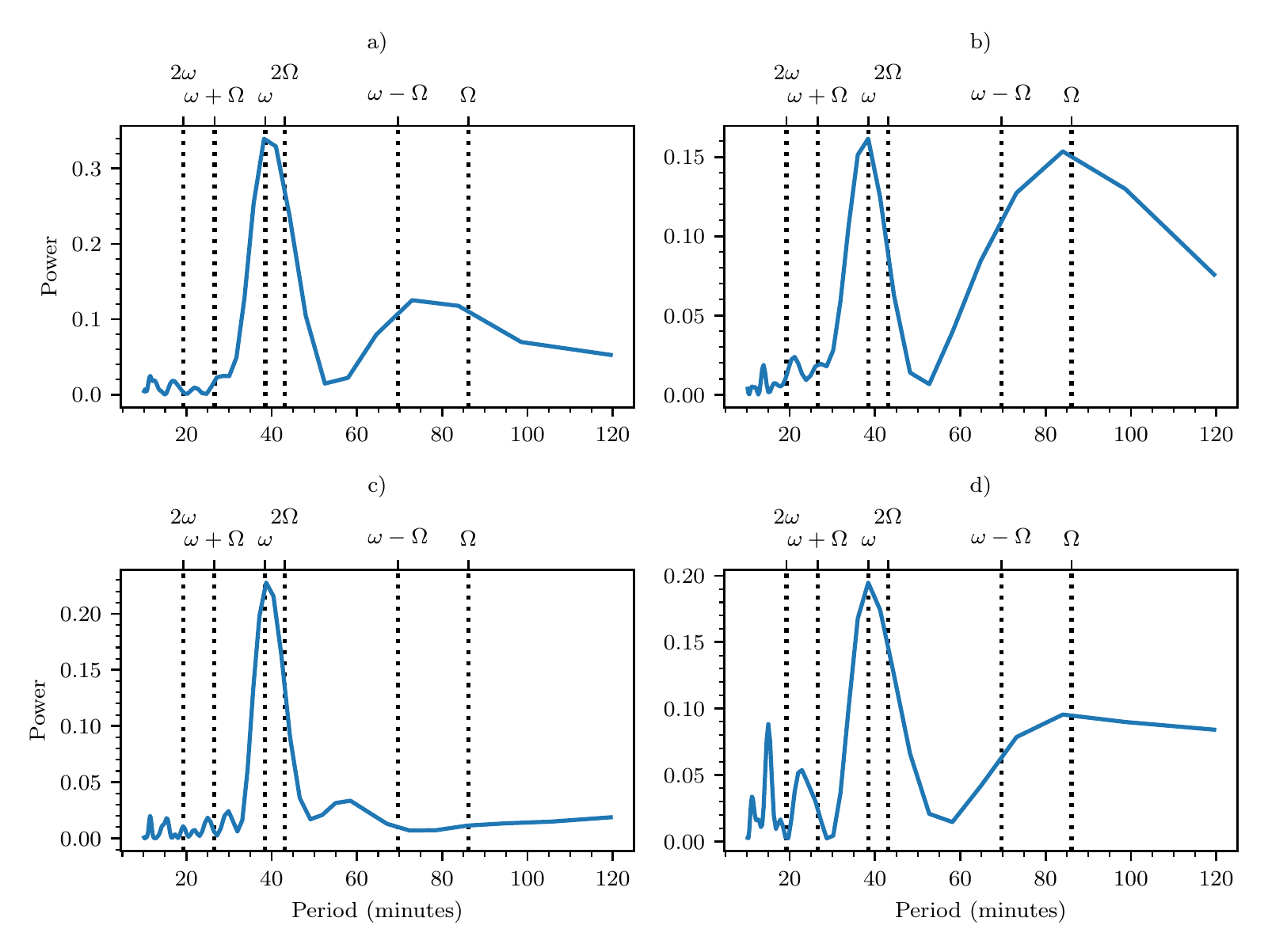}
\caption{Representative periodograms from the analysis of each night with some known periods, harmonics and beat annotated. Annotations are presented in terms of frequency as the periodogram is calculated in frequency-space before being converted for ease of interpretation. a) Periodogram from HJD=2458838; b) Periodogram from HJD=2458844 this night shows an outburst and a strong $P_{spin}$ signal and a broad signal close to the known $P_{orbit}$ value; c) Periodogram from HJD=2458880; d) Periodogram from HJD=2458910, this night shows an outburst and a strong $P_{spin}$ signal.}
\label{rep_period}
\end{figure*}

As shown in \autoref{NGTS_curve} the night-to-night mean flux does not vary substantially except during an outburst event which is in line with previous observations of DW Cnc in a high state \citep{2008JAVSO..36...60C}. The variation in brightness seen within each of the nights can be attributed to the signature of the spin and orbital period. In addition to this we also see three nights (HJD=2458793, 2458823 and 2458910) where there is appreciable brightening. When converted to magnitudes these events are equivalent to an increase of $\sim1$mag; there are no nearby bright stars within the aperture used for the photometry which means that despite the pixel scale this value is subject to only a low level of dilution. These brightenings appear to occur only for a single night, although without complete data for the night preceding the event on HJD=2458823 we cannot be definitive. This implies that these events last no more than approximately 36 hours; taken together with the magnitude of the increase in brightness we identify these as outbursts from DW Cnc. These outbursts are shorter and with lower amplitude than the outburst identified by \citet{2008JAVSO..36...60C} although in their observations they noted a rapid decline in brightness after onset of $\sim2.25$ mag in $\sim27$ hours. Due to this rapid decline in brightness it is possible that our NGTS observations missed an earlier brighter peak.

\autoref{rep_period} shows four periodograms representative of those from each of the nights which we considered according to the procedure described in \autoref{data_ana}. Each of the periodograms has been annotated with the known orbital and spin periods in addition to the first harmonic of each, the known beat periods, and some other simple beats included for completeness. We show nights HJD=2458838, 2458844, 2458880 and 2458910, the final of these being the final outburst which we observed.

In total we identified the signature of the spin period in 28 nights of observation with a significance greater that $3\sigma$, representing 60\% of the nights we considered to be of sufficient length. As illustrated in \autoref{NGTS_curve} there appears to be no particular dependency on the mean flux of observations and in our ability to identify the spin period. Unsurprisingly however, there does appear to be a link between observation length and our ability to identify the spin period with nights of shorter observation being more prone to lacking a significant signal. In each of the periodograms contained  in \autoref{rep_period} we can see the signature of the spin period, in each case it stands isolated and considerably above the 1\% FAP level. Considering all of the periodograms with significant spin period peaks we find a value for the spin period to be 38.63±0.46 minutes. This is consistent with previous findings (38.58 ± 0.02 min and 38.58377(6) min \citep{2004MNRAS.349..367R,2004PASP..116..516P}, albeit with increased uncertainty when compared to previous work -- this however should not be considered a marker of any change to the spin period, which we would not have expected to see. 

We are unable to provide explanations for not finding the spin period, either by Lomb-Scargle or Gaussian Processes analysis, in some of those nights which we considered to be of sufficient length to contain an identifiable spin period. In most of these nights there is very little variation in the mean flux between nights where we cannot recover the period and immediately proceeding or succeeding nights where we do; similarly the flux ranges in a night show similar consistency. We also investigated the flux error values associated with these observations but this did not yield evidence of any reason as to why the signal could not be recovered, however we do see a loose correlation between the Moon phase and those nights which we did not identify the signal. We can only speculate that for the outburst at HJD=2458823 that part of the general trend of the brightening obscures the spin period signal -- indeed we see a signal $\sim42$ minutes during this night which may have its physical origin in the spin period, however we do not feel confident that this is a signature of the spin period. The report of the previously seen outburst by \citet{2004PASP..116..516P} did not seek evidence of the spin period in data originating from solely the outburst.

Our analysis did not yield considerable evidence for the orbital period through the Lomb-Scargle method, which was somewhat inline with expectations given the relatively weak nature of the signal compared to the spin period. However in four of the nights, incidentally those with some of the longest duration observations, we see a diminished and wide peak at the known value of the orbital period, see \autoref{rep_period} b. Analysis with Gaussian Process regression however, shows the signal of the orbital period in 7 nights; this discrepancy arises from the dominant red noise obscuring this in the Fourier based analysis which the Gaussian Process method is able to take account of. Taken together this provides some limited evidence for the orbital period in these data. We similarly do not find any compelling evidence of any of the known beat periods in these data with either method; although likewise this is inline with our expectations of the data given their known strength in previous work and their period in the context of the duration of observations we considered. 

\section{Discussion}
\subsection{The Spin Period}
The power spectra of IPs gives insight to the accretion process taking place in the binary with the presence and strength of peaks in the spectra indicating different accretion geometries. \citet{1999MNRAS.309..517F} predict power spectra resulting from different accretion states, namely disc and discless (i.e. stream fed). Their model showed that the presence of the spin period in the resulting power spectrum was a signature of an accretion disc. Applying this model to power spectra we found of DW Cnc would suggest that the accretion is predominately disc fed.

The question remains as to the apparent absence of the spin period in the low state discussed by \citet{2020MNRAS.494.4110S}. The power spectra presented in that work (c.f. Figures 9 \& 10) shows the presence of a beat period, $\omega-\Omega$, although they note that power is diminished compared to the high state \citep[e.g][]{2004PASP..116..516P} and that the signature of the orbital period is a factor of 10 stronger. Given that the model of \citet{1999MNRAS.309..517F} requires ``that for stream-fed accretion, significant power is expected in the optical at the frequency [of the beat period]'' and as this power was absent during the low state, we conclude therefore that the low state of DW Cnc is not stream fed, equally the absence of a spin period signal also suggests the absence of disc fed accretion.

\citet{2020MNRAS.494.4110S} proposed two possible solutions as to the apparent absence of the spin period of DW Cnc during the low state; that the drop to the low state and associated reduction in the mass accretion rate inhibited the lighthouse effect which reveals the spin period to us or that in the post outburst state the system is required to be in a low state as predicted by \citet{2017A&A...602A.102H}. This second scenario requires that the outburst be governed by the DIM. This was considered unlikely by the authors at the time due to previous identifications of the spin period made after the 2008 outburst \citep{2008JAVSO..36...60C}. We are confident discounting this scenario also as each of the outbursts we observe are found in a high state, therefore ruling out the DIM, with the spin period seen throughout these observations.

The first scenario whereby the low state is the cause of the apparent loss of the spin period is significantly more likely; when first proposed \citet{2020MNRAS.494.4110S} indicated that if this was the case then the eventual return to a high state would be be coupled with the return of an observable spin period. Our results from this most recent high state clearly indicate the return of the spin period. We conclude that the lack of a signature in the low state is due to mass transfer effectively ceasing.

\subsection{Outbursts}
We have identified three, relatively short, outbursts that have occurred in quick succession, relative to other IPs, with a mean recurrence time of $\sim60$ days. These findings are corroborated by evidence gathered from all sky photometry shown in \autoref{All_Sky}, with a number short lived events, likely to be outbursts, which have a similar amplitude and recur on a similar timescale. This recurrence time is significantly shorter than any previously identified for an IP; EX Hya is perhaps the most prolific outbursting IP with a recurrence time of $\sim1.5$ years for outbursts of duration 0.25-1 day \citep{2000MNRAS.313..703H}. The form taken by the outbursts which we see is consistent with that seen in other IPs that have shown outbursts in (suspected) high states \citep[see][for examples]{1989A&A...219..195V,2000MNRAS.313..703H}, we observe similar outburst amplitude and a duration on the order of 1 d.

From the conclusions of \citet{2017A&A...602A.102H} we believe that the outbursts which we see are not a result of disk instability as it was concluded that these could only occur from IPs in their low state and for each of the outbursts we detect DW Cnc is in a high state. As such we identify these to be outbursts to have one of two possible origins. Magnetic instabilities induced by the interaction of the magnetorotational instability of the accretion disc and the WD magnetic field. The duration of these events seems to support this, the DIM is not able to support outbursts in IPs of less than $\sim1$ d, and the observed outburst duration in our data is consistent with these outbursts being shorter than $\sim1$ d.

The alternative origin is magnetically-gated outbursts similar to those recently identified in TW Pic \citep{2021NatAs.tmp..201S}, which requires that the co-rotation radius be approximately equal to the magnetospheric radius where a barrier to accretion is formed. This is conceivably the case in DW Cnc assuming a magnetic moment $\sim3\times10^{32} $ G cm$^{3}$ and an accretion rate $\sim 10^{15}$ g s$^{-1}$, typical for an IP of DW Cnc's orbital period \citep[][c.f. Figure 11]{2011ApJS..194...28K}, although no reliable magnetic field strength measurements for DW Cnc exist at this time. However, if this is indeed the origin of these outbursts then it has implications for the dominant accretion mode of the high state; the barrier to accretion at the co-rotation radius in DW Cnc would exist at $\sim 0.34R_{\sun}$ \citep{2012MNRAS.420..416D}, assuming a WD mass of $0.75M_{\sun}$ \citep{2011ApJS..194...28K}. However for an accretion disk to form the material must infall to at least the circularisation radius, $\sim0.20R_{\sun}$. As such if it is the case that magnetic gating is present during the high state then it cannot be the case that the accretion is disc fed. Yet, there is a strong detection of the spin signal in our light curves, which is considered a strong indicator for the presence of an accretion disc. Reconciling these differences likely requires high time resolution spectroscopy.

Although magnetic gating is a viable physical origin to the outbursts seen in DW Cnc the discrepancy that it raises in relation to the high state accretion mode leads us to conclude that this is the less likely origin. This is reinforced when comparing the outbursts we have presented with those presented by \citet{2021NatAs.tmp..201S} in TW Pic; the outbursts in DW Cnc occur in the high state as opposed to the low in TW Pic, they are also of longer duration and reoccur less frequently.

The high apparent recurrence rate of these high state outbursts is entirely unprecedented in IPs, being many factors greater than that seen in other IPs before; this finding has been made possible with the high cadence of our observations relative to previous studies. These findings may indicate that outbursts in IPs occur more frequently than current observations suggest and that they outburst at a rate comparable to that seen in non-magnetic dwarf novae. This finding may address the existing uncertainty surrounding the frequency of outbursts in IPs when compared to predictions made considering the accretion physics in other dwarf novae. Confirmation of this will however require concerted efforts from survey telescopes with high cadence and suitable photometric depth in order to make sufficient long term observations available for analysis.

\section{Conclusions}
We have analysed high cadence photometry of DW Cnc gathered over $\sim4$ months by NGTS in order to study its high state behaviour. We searched for the known spin period of DW Cnc which had recently been reported to be absent. We were able to confirm the presence of the spin period signal; which had returned when DW Cnc re-entered a high state. Our finding addresses the question raised by and confirms the prediction of \citet{2020MNRAS.494.4110S}. 

The NGTS data also allowed us to identify three new high state outbursts in DW Cnc each appearing to last less than a day, which expands the total number of times that DW Cnc has been seen to outburst by a factor of four. We believe that these outbursts are of a type that originate either from the instability generated as a result of an interaction between the magnetic field in the WD and the magnetorotational field generated by the accretion disc or via magnetic gating. Either of these origins would make  DW Cnc quite different to those seen in much of the rest of the IP population. These outbursts were seen with an unprecedented recurrence time ($\sim60$ d) for IPs which opens the door to re-examining these systems with similar high cadence observations in relation to their outbursting behaviour in order to identify other such short lived and relatively frequent outbursts.

\section*{Acknowledgements}
This research is based on data collected under the NGTS project at the ESO La Silla Paranal Observatory. The NGTS facility is funded by a consortium of institutes consisting of the University of Warwick, the University of Leicester, Queen's University Belfast, the University of Geneva, the Deutsches Zentrum für Luftund Raumfahrt e.V. (DLR; under the `Gro{\ss}investition GI-NGTS'), the University of Cambridge, together with the UK Science and Technology Facilities Council (STFC; project reference ST/M001962/1).

The Gravitational-wave Optical Transient Observer (GOTO) project acknowledges the support of the Monash-Warwick Alliance; University of Warwick; Monash University; University of Sheffield; University of Leicester; Armagh Observatory \& Planetarium; the National Astronomical Research Institute of Thailand (NARIT); Instituto de Astrof\'{i}sica de Canarias (IAC); University of Portsmouth; University of Turku, and the UK Science and Technology Facilities Council (STFC grant numbers ST/T007184/1, ST/T003103/1).

Parts of this research were conducted by the Australian Research Council Centre of Excellence for Gravitational Wave Discovery (OzGrav), through project number CE170100004.

Armagh Observatory \& Planetarium is core funded by the Northern Ireland Executive through the Department for Communities. C. Duffy acknowledges STFC for the receipt of a postgraduate studentship.

M. R. K. acknowledges support from the Irish Research Council in the form of a Government of Ireland Postdoctoral Fellowship (GOIPD/2021/670: Invisible Monsters).

Based on observations obtained with the Samuel Oschin 48-inch Telescope at the Palomar Observatory as part of the Zwicky Transient Facility project. ZTF is supported by the National Science Foundation under Grant No. AST-1440341 and a collaboration including Caltech, IPAC, the Weizmann Institute for Science, the Oskar Klein Center at Stockholm University, the University of Maryland, the University of Washington, Deutsches Elektronen-Synchrotron and Humboldt University, Los Alamos National Laboratories, the TANGO Consortium of Taiwan, the University of Wisconsin at Milwaukee, and Lawrence Berkeley National Laboratories. Operations are conducted by COO, IPAC, and UW. We thank Jan van Roestel for discussions regarding the ZTF data.

We thank the reviwer, J.~P. Lasota, for their helpful and insightful comments. 

\section*{Data Availability}
NGTS data used in this work is available as part of the online supplementary material.  

ZTF data products used in this work are publicly availble at \url{https://irsa.ipac.caltech.edu/Missions/ztf.html} and ASAS-SN data products are available at \url{https://asas-sn.osu.edu/photometry}.

GOTO data products will be available as part of planned GOTO public data releases.

\bibliographystyle{mnras}
\bibliography{References.bib}

\bsp	
\label{lastpage}
\end{document}